\documentclass[twocolumn]{article}

\usepackage[utf8]{inputenc}
\usepackage[a4paper, margin=1.8cm, columnsep=0.6cm]{geometry}
\usepackage{authblk}
\usepackage{setspace}
\usepackage{graphicx}
\graphicspath{{./figures/}}
\usepackage{subcaption}
\usepackage{amsmath}
\usepackage{booktabs}
\usepackage{hyperref}
\usepackage{tikz}
\usepackage{amssymb}
\usepackage{subcaption}
\usepackage{braket}

\usepackage[switch]{lineno}       
\usepackage[numbers,sort&compress]{natbib}

\title{\textbf{Implementation of quantum gates by Floquet analysis of kicked quantum system}}

\author[1,2,*]{Andrea De Luca}
\author[3]{Carola Ciarameletti}
\author[1]{Simone Paganelli}

\affil[1]{Dipartimento di Scienze Fisiche e Chimiche, Università dell'Aquila, Coppito-L'Aquila, Italy}
\affil[2]{INFN, Laboratori Nazionali del Gran Sasso, Via G. Acitelli 22, 67100 Assergi (AQ), Italy}
\affil[3]{Dipartimento di Ingegneria e Scienze dell'Informazione e Matematica, Università dell'Aquila, Coppito-L'Aquila, Italy}
\affil[*]{andrea.deluca4@graduate.univaq.it}

\date{\today}

\begin{document}
\twocolumn[
\begin{@twocolumnfalse}

\maketitle

\begin{abstract}
	Precise control of multi-qubit architectures remains a critical bottleneck in superconducting quantum processors. In this work, we investigate the synthesis of high-fidelity quantum operations and state transfer protocols within an extended superconducting linear chain, scaling from a foundational three-qubit processor up to a seven-site network. Using Floquet theory, we initially model the periodic external drive as a train of delta-like pulses, providing a rigorous stroboscopic description where the quantum control problem is mapped onto the resonance conditions of the quasi-energy spectrum. Through a combination of the Baker-Campbell-Hausdorff expansion and Floquet spectral decomposition, we analytically identify the optimal driving parameters, which are subsequently refined using the Covariance Matrix Adaptation Evolution Strategy (CMA-ES) to efficiently navigate the highly irregular fidelity landscape. In the three-qubit architecture, this approach enables the high-fidelity synthesis of the iSWAP gate. Moving toward experimental feasibility, we extend the global driving protocol to a seven-site chain by implementing physically realistic periodic trains of finite-width Gaussian pulses. We demonstrate the high-fidelity activation of distinct double-excitation transport channels with ultra-short gate durations $t_\text{gate}$ (around $170\text{ ns}$), achieving a clear scale separation from the energy-relaxation times ($T_1$) typical of fixed-frequency transmon devices with tunable couplers, of the kind realized on IBM Quantum hardware \cite{architecture, roth2017analysis,jurcevic2021demonstration, sung2021realization}; since the argument depends only on the dimensionless ratio $t_{\text{gate}}/T_1$, the conclusion extends naturally to platforms with different coherence times. Finally, we benchmark the protocol's stability under realistic hardware imperfections, revealing a heightened sensitivity to static parameter disorder at the sub-percent level ($\eta \sim 10^{-3}$) driven by spectral crowding, and we discuss how closed-loop ring topologies could mitigate this constraint. This comprehensive framework bridges time-periodic control theory and the practical requirements of robust, scalable quantum gate engineering.
\end{abstract}
\vspace{1em}
\end{@twocolumnfalse}
]


\section{Introduction}

Quantum computing has attracted significant interest due to its potential for solving complex computational problems, such as in cryptography \cite{Quantum_cryptography}, combinatorial optimisation and quantum chemistry \cite{Quantum_chemistry}. Unlike classical architecture, in which information is confined to a binary value (bit), in a quantum processor information is encoded in a qubit, which takes on values on the Bloch sphere. Quantum superposition and entanglement between qubits make it possible to significantly speed up the execution of various classes of algorithms \cite{benenti2004principles}.

The basic mechanism of a quantum computer relies on the implementation of high-fidelity unitary transformations within quantum circuits \cite{Williams2011}. The aim is to create gates with sufficient fidelity to negate intrinsic circuit errors, such as stochastic fluctuations or systematic errors that could lead to the rapid loss of quantum coherence. While single-qubit operations in superconducting circuits have achieved fidelities surpassing $99.99\%$\cite{single_qubits}, multi-qubit gates, essential for universal quantum computation, typically exhibit higher error rates, with the best implementations exhibiting a high fidelity \cite{multi_qubits}. The demonstration of quantum computational advantage \cite{arute2019quantum} has increased interest in scalable, high-fidelity multi-qubit control protocols.

Various physical systems have been developed to overcome these issues, such as trapped ions \cite{trappedions}, semiconductor quantum dots \cite{semiconducting_qubits}, Rydberg atoms \cite{Rydberg}, and photonic systems \cite{photonic_system}. In the domain of superconducting qubits \cite{superconducting_qubits}, architecture-specific interactions can be engineered using tunable couplers. The transmon qubit \cite{koch2007charge} is a key element in the construction of new superconducting processors, thanks to its low sensitivity to charge noise. Specifically, Refs. \cite{architecture, roth2017analysis} demonstrate an architecture where two fixed-frequency transmon qubits are coupled via a flux-modulated ancilla transmon acting as a tunable bus. By dynamically tuning the bus frequency, it is possible to activate resonant exchange interactions, including iSWAP and two-photon processes \cite{poletto2012entanglement}, while maintaining a dispersive regime during idle states. Further refinements of tunable coupling schemes \cite{yan2018tunable} and analytical modelling of parametric modulation \cite{didier2018analytical} have enabled the development of a solid theoretical foundation for high-fidelity gate engineering in these architectures.

In this work, we investigate the role of Floquet dynamics as a framework for controlling such periodically driven systems. Floquet theory \cite{bukov2015universal, sambe1973steady} allows the mapping of a time-periodic Hamiltonian into an effective time-independent description, providing a rigorous method for engineering the system's interaction channels. The quasi-energy spectrum, first systematically studied by Shirley \cite{resonance} and Sambe \cite{sambe1973steady} in the context of periodically driven quantum systems, dictates the resonance dynamics: resonant transitions occur whenever quasi-energy differences match an integer multiple of the driving frequency \cite{resonance}. We focus on a specific class of modulation known as \textit{kicked systems} \cite{kicked}, where the driving is modelled as a periodic sequence of delta-like pulses. These models allow for an analytically tractable limit that reveals resonance conditions and frequency renormalization through a perturbative expansion \cite{baker}.

The paper is organized as follows. In Sec.~\ref{sec:three-qubit}, we apply the Floquet analysis to the three-qubit processor of Ref. \cite{roth2017analysis} under ideal delta-like modulation. We optimize the kick strength $\lambda$ and driving period $T$ through a combination of analytical resonance estimates and the Covariance Matrix Adaptation Evolution Strategy (CMA-ES) \cite{hansen2016cma}, a highly effective derivative-free evolutionary algorithm for navigating the highly irregular fidelity landscape near Floquet resonances, demonstrating the high-fidelity synthesis of the iSWAP gate \cite{iSWAP}. 

In Sec.~\ref{sec: Scaling 7}, we depart from idealized, infinitely short driving profiles by implementing a physically realistic periodic train of Gaussian pulses. We explore the scalability of this global Floquet engineering approach within a larger seven-site superconducting linear chain, specifically co-optimizing complex multi-excitation transfer processes. Finally, in Sec.~\ref{sec:robustness}, we provide a robustness analysis of the protocol. We benchmark its performance under realistic open-system timescales and subject the network parameters to static hardware disorder and quantum projection noise, mapping out the operational boundaries, structural vulnerabilities, and fundamental spectral limitations of purely linear topologies.

\section{Three-qubit system setup and Floquet Driving} \label{sec:Floquet_theory}
To implement the gate synthesis protocol, we model the periodic external drive as a train of delta-like pulses (a kicked system) \cite{kicked}. This choice provides a mathematically rigorous yet tractable framework for Floquet engineering \cite{oka2019floquet, eckardt2017colloquium, holthaus2016floquet}. The time-dependent Hamiltonian of the driven system is given by:
 
 \begin{equation}
 	\widehat{H}(t) = \widehat{H}_0 + \lambda\widehat{V}\sum_{n=-\infty}^\infty \delta (t-nT).
 	\label{delta_train}
 \end{equation}
  Throughout the analytical derivations we set $\hbar = 1$ for convenience, whereas physical quantities in numerical simulations and figures are restored to SI units. Here $\lambda$ denotes the kick strength, and the sharp separation between the free evolution $\widehat{H}_0$ and the instantaneous kick $\widehat{V}$ allows for an exact factorization of the Floquet propagator. By adopting a gauge in which the kick coincides with the start of the period, one obtains:

 \begin{equation}
 	\widehat{U}(t_0+T,t_0)=e^{-i\widehat{H}_0T}e^{-i\lambda\widehat{V}} = e^{-i\widehat{H}_FT},
 \end{equation}
 so that the effective Floquet Hamiltonian is:
 
\begin{equation}
	\widehat{H}_F=-\frac{i}{T}\ln\biggl[e^{-i\widehat{H}_0T}e^{-i\lambda\widehat{V}}\biggr].
\end{equation}

When $[\widehat{H}_0,\widehat{V}] = 0$ this reduces trivially to $\widehat{H}_F = \widehat{H}_0 + \lambda / T \widehat{V}$. While in the non-commuting case the Baker-Campbell-Hausdorff expansion \cite{baker} produces:

\begin{equation}
	\widehat{H}_F = \widehat{H}_0 + \frac{\lambda}{T}\cdot\widehat{V} - \frac{i\lambda}{2}[\widehat{H}_0,\widehat{V}] + \cdots .
\end{equation}
The higher-order commutators in this expansion encode non-trivial effective interactions that would be absent in the static system. This mechanism, referred to as Floquet engineering \cite{oka2019floquet}, enables the selective activation of desired coupling channels through the tuning of $\lambda$ and $T$, and that will form the operational principle underlying the gate synthesis protocol developed in the following sections.

\subsection{Three-qubit system setup} \label{sec:three-qubit}

We consider a superconducting circuit architecture consisting of two fixed-frequency transmon qubits ($Q_1$ and $Q_2$) coupled via a flux-tunable transmon bus ($TB$), as schematically shown in Fig. \ref{fig:tree_qubits_system}. The transmon design \cite{koch2007charge, superconducting_qubits}, because of its reduced sensitivity to charge noise, is  widely used in modern superconducting processors. The system is described by the multi-qubit Hamiltonian:

\begin{figure}[b]
	\centering
	\includegraphics[width=0.4\textwidth]{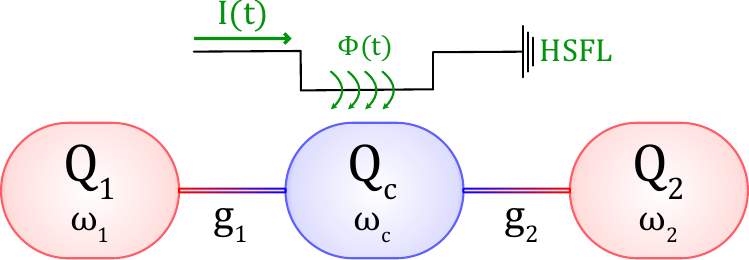}
	\caption{\begin{small}
			Circuit scheme taken from \cite{roth2017analysis} of the device which consist of two fixed-frequency transmons ($Q_1$,$Q_2$) capacitively coupled to a flux tunable transmon ($TB$). The tunable coupler is controlled by a high-speed flux line (HSFL) providing a current $I(t)$ and a consequent flux $\Phi(t)$.
	\end{small}}
	\label{fig:tree_qubits_system}
\end{figure}

\begin{equation}
	\widehat{H} = - \sum_{i=1,2} \frac{\omega_i}{2}\widehat{\sigma}_i^z - \frac{\omega_c(t)}{2}\widehat{\sigma}_c^z + g_1\widehat{\sigma}_1^x\widehat{\sigma}_c^x + g_2\widehat{\sigma}_2^x\widehat{\sigma}_c^x,
\end{equation}
where $\widehat{\sigma}_i^\alpha$ are the Pauli operators for the $i$-th qubit and the coupler ($c$). The computational qubits $Q_1$ and $Q_2$ are characterized by fixed frequencies $\omega_i$, while the central coupler mediates the interaction through a capacitive $XX$-type coupling of strength $g_i$ \cite{roth2017analysis, architecture, chen2014qubit}. The coupler frequency $\omega_c(t)$ is modulated by external time-dependent magnetic flux $\Phi(t)$ according to:

\begin{equation}
	\omega_c(t) = \omega_c^0\sqrt{\biggl|\cos\biggl(\frac{\pi \Phi (t)}{\Phi_0}\biggr)\biggr|},
\end{equation}
where $\omega_c^0$ is the maximum frequency at zero flux and $\Phi_0$ is the flux quantum ($h/2e$). This tunability allows one to activate entangling operations such as iSWAP and bSWAP gates \cite{iSWAP}, and has been exploited in tunable coupling schemes of increasing complexity \cite{yan2018tunable, didier2018analytical, poletto2012entanglement, mundada2019suppression, strand2013first}.

\subsection{The pulsed Driving Approximation and stroboscopic evolution}

The external driving is modelled as a periodic sequence of Gaussian pulses acting on the longitudinal degree of freedom of the coupler: 

\begin{equation}
	\widehat{H}(t) = \widehat{H}_0 + \widehat{H}_{\text{int}} + \sum_{n} A\cdot e^{-\frac{(t-nT)^2}{2\sigma^2}}\widehat{\sigma}_c^z,
\end{equation}
where $\widehat{H}_0 = -\sum_{i=1,2} \frac{\omega_i}{2}\widehat{\sigma}_i^z - \frac{\omega_c^0}{2}\widehat{\sigma}_c^z$ is the bare single-qubit Hamiltonian, and $\widehat{H}_{\text{int}} = g_1\widehat{\sigma}_1^x\widehat{\sigma}_c^x + g_2\widehat{\sigma}_2^x\widehat{\sigma}_c^x$ accounts for the static capacitive interaction between the transmons and the central bus.

In the impulsive limit $\sigma \ll T$, this reduces to the kicked-system framework introduced in Sec \ref{sec:Floquet_theory}, with the modulation described by a periodic delta-pulse train acting on the coupler:

\begin{equation}
	\widehat{H}(t) = \widehat{H}_0 + \widehat{H}_{\text{int}} + \lambda\sum_{n=0}^{\infty}\delta(t-nT)\widehat{\sigma}_c^z,
\end{equation} 
where $\lambda = A\sqrt{2\pi}\sigma$ is the integrated kick strength. Adopting the Floquet gauge defined in Appendix \ref{app:floquet_foundations}, in which each kick coincides with the start of the driving period, the $n-$period stroboscopic evolution operator can be written as:

\begin{equation}
	\widehat{U}(t_0+nT,t_0) = \biggl[e^{-i(\widehat{H}_0+\widehat{H}_{\text{int}})T}e^{-i\lambda\widehat{\sigma}_c^z}\biggr]^n, 
	\label{time_operator} 
\end{equation}
in this regime, the bus mediates transitions between the fixed-frequency qubits through stroboscopic exchange interactions, and the gate synthesis problem is limited to identifying the driving parameters $(\lambda, T)$ that maximize the transition fidelity:
\begin{equation}
	\mathcal{F}_{i\longrightarrow f} = |\langle f|\widehat{U}(nT)|i\rangle |^2,
	\label{Fidelity_def}
\end{equation}
where $|i\rangle$ and $|f\rangle$ denote the initial and final computational states.

\subsection{Resonance conditions and analytical estimates} \label{subsec: resonance condition}

The periodic delta-kick train acts exclusively on the bus degree of freedom, modifying the quasi-energy spectrum of the full system. This shift in the spectrum can be seen as a renormalization of the effective coupler frequency: the drive induces a phenomenological shift $\omega_{\text{eff}}(\lambda,T)$ that depends explicitly on the driving parameters, but does not constitute a fundamental parameter of the theory. Thus, one should ask under which conditions on ($\lambda,T$) this modified quasi-energy structure supports a resonant transition between two selected computational states $|i\rangle$ and $|f\rangle$.
Within the Floquet framework, the quasi-energies $\epsilon_\alpha$ are defined modulo the driving frequency $\Omega = 2\pi/T$, in direct analogy with quasi-momenta defined modulo a reciprocal lattice vector in spatially periodic systems \cite{condensed_matter}. As a consequence, resonance is not restricted to a simple matching between $\omega_{\text{eff}}$ and the bare energy difference $\Delta E = \omega_f-\omega_i$. Rather, a controlled transition between $|i\rangle$ and $|f\rangle$ is induced whenever the quasi-energy difference satisfies the Floquet resonance condition of order $m$ \cite{resonance, sambe1973steady}:

\begin{equation}
	\Delta \epsilon_{f,i} = m\Omega, \hspace{0.3cm} m\in \mathbb{Z}
	\label{floquet_resonance}
\end{equation}

This condition shows that quasi-energies are defined modulo $\Omega$: transitions can occur through the exchange of $m$ energy quanta $\Omega$ between the system and the periodic drive, a mechanism with no analogue in time-independent Hamiltonians. The full derivation of this condition from the periodicity of the Floquet spectrum is given in Appendix \ref{app:Resonance}.

To use this condition to create constraints on the ($\lambda,T$), we perform a spectral decomposition of the fidelity in the Floquet eigenbasis $\{|\alpha\rangle\}$. Expanding the computational states as:

\begin{equation}
	|i\rangle = \sum_\alpha c_\alpha|\alpha\rangle \qquad |f\rangle = \sum_\alpha b_\alpha |\alpha\rangle,
\end{equation}

and expressing the stroboscopic evolution operator in the Floquet basis, the fidelity takes the form: 

\begin{equation}
\mathcal{F}_{i \to f} = \sum_{\alpha} |c_\alpha|^2 |b_\alpha|^2 + \sum_{\alpha \neq \beta} c_\alpha b_\alpha^* c_\beta^* b_\beta e^{-i(\epsilon_\alpha - \epsilon_\beta)nT}.
\label{resonance}
\end{equation}

The first term is a time-independent background determined by the static overlaps between the computational states and the Floquet modes. The second term consists of oscillatory interference contributions whose phases accumulate with each stroboscopic step. Resonant population transfer toward $|f\rangle$ occurs when these interference terms add constructively over successive kicks, driving $\mathcal{F}_{i\to f}$ sustainedly toward unity. This requires precisely that the dominant quasi-energy differences satisfy $\Delta\epsilon_{\alpha\beta}=m\Omega$, recovering the resonance condition above. In practice, only those pairs $|\alpha\rangle, |\beta\rangle$ carrying the largest weights $c_\alpha b_\alpha^*$, i.e. those Floquet modes with maximal overlap onto the initial and final computational states, contribute significantly to the dynamics; the remaining terms average out stroboscopically. The Fourier spectrum of the fidelity time series therefore directly probes the resonant quasi-energy structure of the driven system, which will be used in Sec \ref{sec:quantum_gate} for spectral validation of the gate performance.
The complete derivation is reported in Appendix \ref{app:estimate}, and the resulting estimates are used to initialize for the numerical optimization described in Sec. \ref{subsec:computational implementation}.

\subsection{Numerical optimization via CMA-ES}\label{subsec:computational implementation}

While the analytical analysis in Sec \ref{subsec: resonance condition} identify the resonance regions, the non-commuting nature of $\widehat{H}_0$ and $\widehat{V}$, combined with the discrete stroboscopic structure of $nT$ require a fully numerical treatment. We implement a computational pipeline in Python to maximize $\mathcal{F}_{i \to f}$ over the parameter space $(\lambda, T, n)$.

For any given pair ($\lambda,T$), the one-period propagator is constructed via matrix exponentiation (\texttt{scipy.linalg.expm}) and iterated stroboscopically. To account for physical coherence time constraints, the objective function is defined as the peak fidelity within a window of $N_\text{max} = 250$ kicks:

\begin{equation}
	\mathcal{F}_{\text{max}}(\lambda, T) = \max_{n\in \{1,\ldots,N_{\text{max}}\}} |\langle f|\widehat{U}_T^n|i\rangle|^2,
\end{equation}
reducing the problem to a two-dimensional search over the continuous variables $\lambda$ and $T$

To explore the resulting landscape, which exhibits sharp, non-convex peaks near the Floquet resonance of Eq. \eqref{floquet_resonance}, we employ the \textbf{Covariance Matrix Adaptation Evolution Strategy} (CMA-ES) \cite{hansen2016cma}, an effective derivative-free evolutionary algorithm for high-frequency, non-convex optimization problems. Such derivative-free and adaptive optimization techniques are widely adopted for automated gate calibration and optimal quantum control in superconducting architectures \cite{kelly2014optimal, werninghaus2021leakage}. The procedure consists of four stages:

\begin{itemize}
	\item \textbf{Initialization:} The search is seeded in the vicinity of the analytically estimated resonance value to accelerate convergence;
	\item \textbf{Stochastic Sampling:} a population of candidates $(\lambda, T)$ is drawn from a multivariate normal distribution;
	\item \textbf{Covariance Matrix Adaptation:} the distribution mean and covariance are updated at each iteration, progressively learning the local topology of the fidelity landscape;
	\item \textbf{Extraction of optimal $n$:} Once ($\lambda_{opt}, T_{opt}$) are identified, the integer kick number $n_{\text{best}}$ that achieves peak fidelity is extracted, defining the total gate duration.
\end{itemize}

This procedure bypasses the issues associated with truncated BCH expansions and identifies high-fidelity operating points for the iSWAP and bSWAP gate implementations presented in Sec \ref{sec:quantum_gate}.

\section{Floquet-Induced State Transfer in the Three-Qubit System} \label{sec:quantum_gate}

We now apply the framework developed in Sec. \ref{sec:three-qubit} to implement quantum gates in the three-qubit architecture. We first compare the driven system with the undriven case, then present the optimized iSWAP gate realization, and analyze the resonance mechanism responsible for the transfer through spectral analysis of the fidelity dynamics.

All simulations use the device parameters reported in Table \ref{tab:parameters}, expressed in natural units consistent with the GHz-scale regime typical of superconducting transmon processors \cite{koch2007charge,roth2017analysis}. The computational basis states are denoted as $|\alpha_1,\alpha_2,\alpha_c\rangle$, where the first two indices refer to the fixed-frequency qubits and the third to the tunable bus.

\begin{table}[h]
	\centering
	\begin{tabular}{l l l}
		\hline
		\textbf{Parameter} & \textbf{Value $(GHz)$} & \textbf{Description} \\
		\hline
		$\omega_1, \omega_2$ & $4.4,4.6$ & Fixed-frequency qubits \\
		$\omega_c$ & $4.5$ & Idle coupler frequency \\
		$g_1,g_2$ & $0.400,0.401$ & Coupling strengths \\
		\hline
	\end{tabular}
	\caption{\begin{small}
			System parameters used for numerical simulations.
	\end{small}}
	\label{tab:parameters}
\end{table}

\subsection{iSWAP Gate Synthesis}

The target process is the iSWAP gate, corresponding to the coherent transfer of a single excitation between the computational qubits via the transition $|100\rangle \to |010\rangle$. This process is particularly suitable for evaluating the Floquet driving protocol, as it requires the system to favour a single exchange channel while suppressing the others \cite{iSWAP, dewes2012quantum}.

\begin{figure}[b!]
	\centering
	\includegraphics[width=0.45\textwidth]{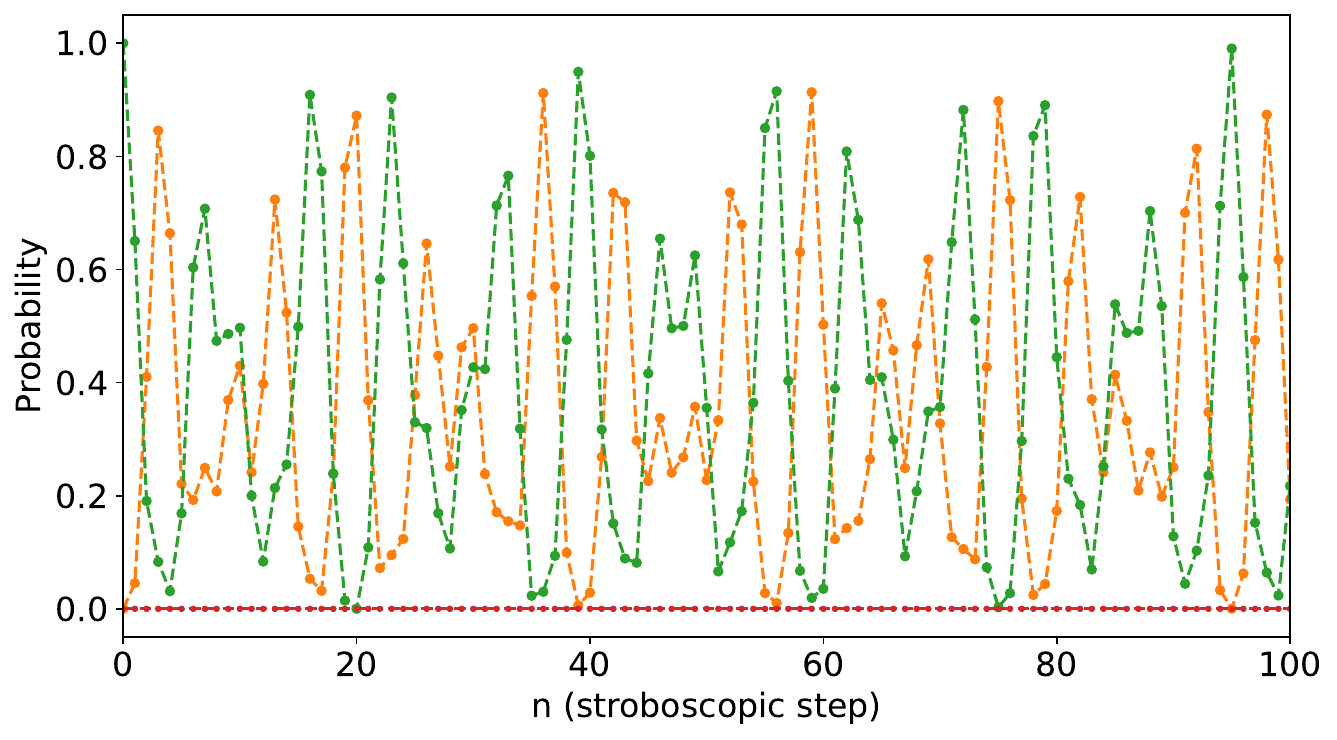}
	\caption{\begin{small}
			\textbf{Unmodulated system dynamics.} Stroboscopic evolution of the state populations for $\lambda=0$, where the $y$-axis denotes the probability of measuring the system in a given basis state at step $n$. The initial state $|100\rangle$ (green dashed line) and the target state $|010\rangle$ (orange dashed line) exhibit disordered oscillations, resulting in poor population transfer fidelity in the absence of Floquet resonance engineering.
	\end{small}}
	\label{no_kick_evolution}
\end{figure}

For comparison, we first consider the undriven system ($\lambda=0$). In the absence of periodic modulation, the static $XX$ couplings produce a complex hybridization among energetically proximate states, and the excitation fails to follow a selective transfer channel. The resulting dynamics exhibit irregular oscillations with substantial leakage into unwanted configurations, as shown in Fig. \ref{no_kick_evolution}.

\begin{figure}[t!]
	\centering
		\includegraphics[width=0.45\textwidth]{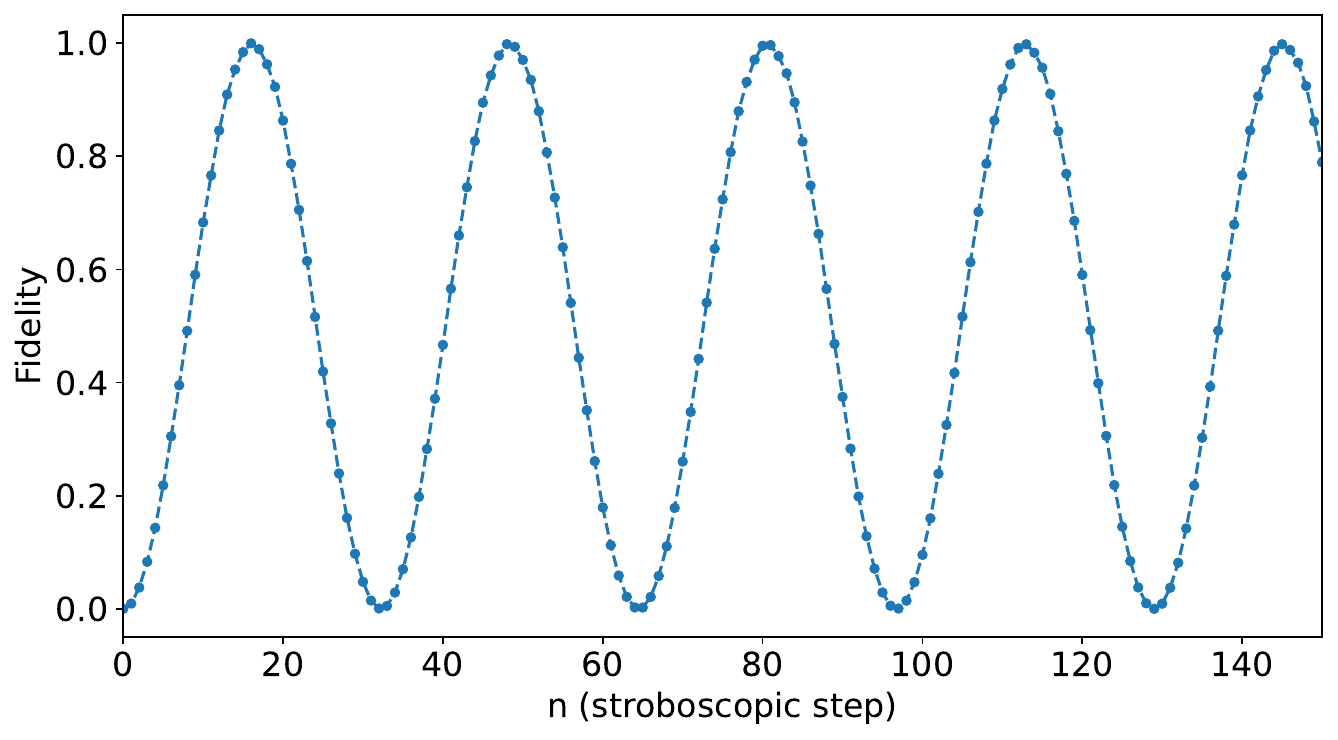}
	\caption{\begin{small}
			\textbf{Optimized iSWAP evolution.} Fidelity evolution [Eq.\eqref{Fidelity_def}] for the state transfer from the initial state $|100\rangle$ to the target state $|010\rangle$ under optimized Floquet driving with parameter $\lambda = 4.62 \hbar$ and $T=10.95 ns$. The Fidelity reaches a maximum close to unity at the optimal kick number $n_\text{best}=16$.
	\end{small}}
	\label{iswap_evolution}
\end{figure}

The introduction of periodic delta-kick modulation changes this behaviour. Optimized driving parameters $(\lambda, T)$ identified via the CMA-ES procedure of Sec. \ref{subsec:computational implementation} \cite{hansen2016cma}, and initialized from the analytical resonance estimates of Sec. \ref{subsec: resonance condition}, produce a coherent transfer between the selected computational states. Specifically, the optimization yields a kick strength $\lambda = 4.62\hbar$ and a driving period $T = 10.95$ $ns$, forcing the effective Hamiltonian to favour the specific transition while strongly suppressing leakage and parasitic couplings. As shown in Fig. \ref{iswap_evolution}, the stroboscopic evolution shows a population transfer between the selected computational states, with a fidelity close to unity at the optimal kick number $n_{\text{best}} = 16$.

\subsection{Spectral Identification of Driving Resonances}

To verify that the observed high-fidelity transfer arises from Floquet resonances rather than numerical artefacts, we analyse the fidelity dynamics in the frequency domain. As shown in Sec \ref{subsec: resonance condition}, the fidelity can be expressed as an interference sum over quasi-energy differences, whose Fourier spectrum therefore shows the resonant structure of the driven system \cite{sambe1973steady, bukov2015universal}.

\begin{figure}[t]
	\centering
	\includegraphics[width=0.45\textwidth]{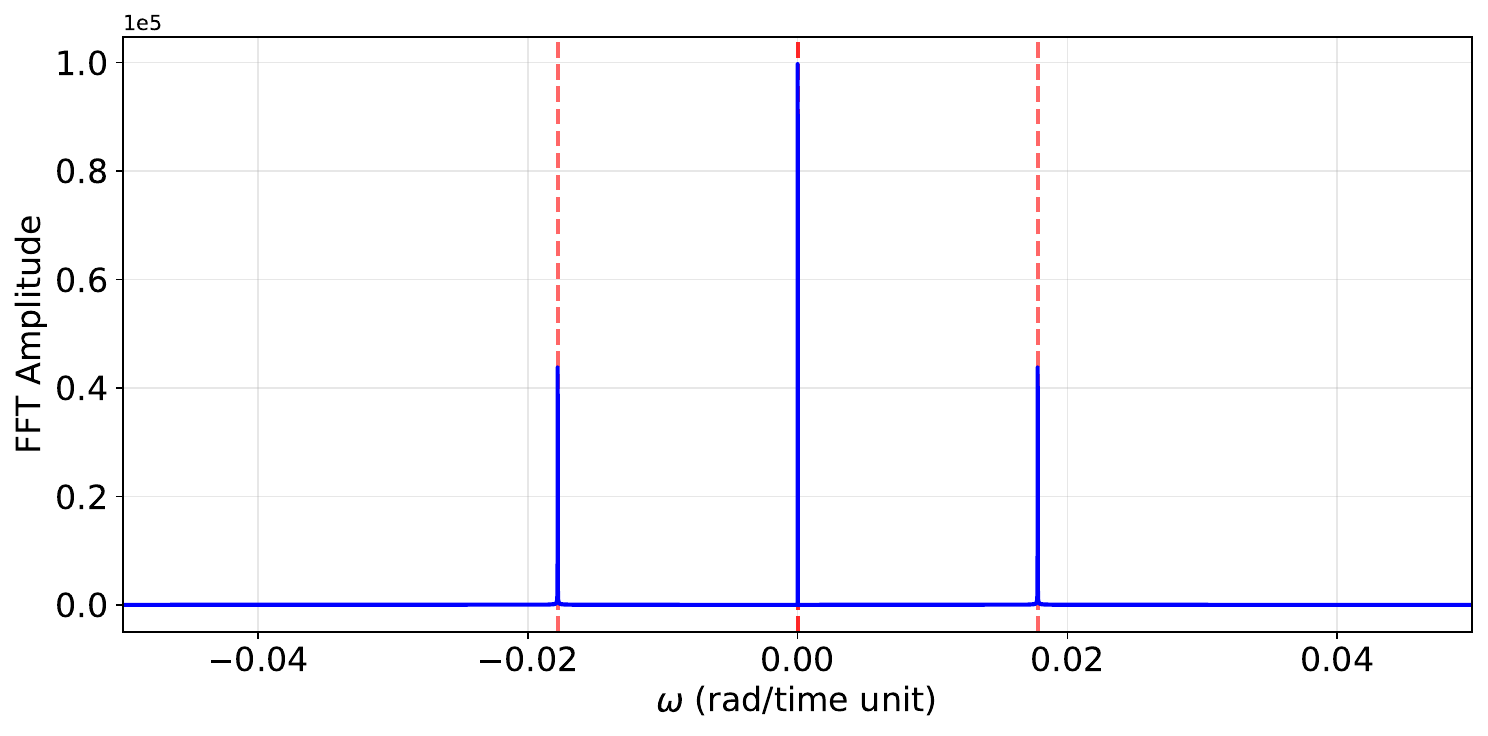}
	\caption{\begin{small}
			\textbf{Spectral analysis of gate fidelity.} Numerical FFT (Fast Fourier Transform) peaks align with analytical quasi-energy transition frequencies under Gaussian-pulsed driving.
	\end{small}}
	\label{fig:fourier_analysis}
\end{figure}

Figure \ref{fig:fourier_analysis} presents the FFT spectrum of the fidelity time series for the optimized iSWAP process. The spectrum exhibits sharp, symmetric peaks whose positions align with the weighted quasi-energy transitions predicted analytically by the effective Floquet Hamiltonian, in agreement with the gate performance being governed by isolated resonances of order $m$, satisfying $\delta\epsilon_{f,i} = m\Omega$. \cite{resonance} The suppression of sidebands and competing resonances shows that the smooth Gaussian-pulse train protocol effectively minimizes crosstalk and leakage despite the finite width of the driving kicks \cite{motzoi2009simple, mundada2019suppression}.

These results show that the proposed protocol can generate high-fidelity iSWAP dynamics in the three-qubit architecture. In Sec \ref{sec: Scaling 7} we investigate whether the same global driving protocol can be extended to larger architectures, where a denser quasi-energy spectrum and multi-excitation processes pose significantly more demanding challenges \cite{altman2015universal}.

\section{Scaling to Linear Chain Architectures: Global Floquet Control} \label{sec: Scaling 7}

We now extend the protocol to a seven-site linear chain composed of four computational qubits ($Q_1,\ldots, Q_4$) interconnected by three tunable couplers ($Q_{C_1}, Q_{C_2}, Q_{C_3} $) interleaved between them:

\begin{equation*}
	Q_1 - Q_{C_1} - Q_2 - Q_{C_2} - Q_3 - Q_{C_3} - Q_4
\end{equation*}
the time-dependent Hamiltonian of the extended system reads:

\begin{equation} 
	\widehat{H}(t) = \sum_{i=1}^{7} \frac{\omega_i}{2}\widehat{\sigma}_i^z + \sum_{i=1}^{6} g_i\widehat{\sigma}_i^x\widehat{\sigma}_{i+1}^x + A \sum_{n, i \in {2,4,6}} e^{-\frac{(t-nT)^2}{2\sigma^2}} \widehat{\sigma}_{i}^z,
	\label{7_qubit_hamiltonian}
\end{equation}
where the static network parameters include the on-site frequencies $\omega_i$ and the fixed inter-qubit couplings $g_i$. To make the model physically realistic, the idealized delta-kicks are replaced by a periodic train of Gaussian pulses with standard deviation $\sigma$, period $T$, and driving amplitude $A$ \cite{motzoi2009simple, didier2018analytical}. All tunable couplers (positioned at index $i \in \{2,4,6\}$) are driven globally with the same stroboscopic control parameters $(A, T, \sigma)$, while the local qubit and coupler frequencies $\omega_i$ are co-optimized alongside $(A, T)$ to navigate more efficiently through the energy landscape, which is denser than that of the three-qubit architecture. 

The computational basis states $|\alpha_1\alpha_2\alpha_3\alpha_4\rangle$ refer exclusively to the four fixed-frequency qubits; the bus degrees of freedom are not tracked in the state labelling, as they act solely as mediators of the inter-qubit interactions.

Unlike the three-qubits case, the seven-qubits case has a larger Hilbert space and, consequently, a more complex resonance structure: multiple competing transition channels become energetically proximate, and the risk of leakage into parasitic manifolds increases considerably. As a result, the process of determining the parameters becomes computationally more demanding and complex, making the CMA-ES strategy of Sec. \ref{subsec:computational implementation} very effective \cite{hansen2016cma}, given that a brute-force search for the parameters would be prohibitively expensive\cite{altman2015universal}.

To assess the protocol across distinct physical processes, we consider three representative double-excitation transitions of increasing spatial complexity:

\begin{itemize}
	\item $|1010\rangle \rightarrow |0101\rangle$
	\item $|1001\rangle \rightarrow |0110\rangle$
	\item $|1100\rangle \rightarrow |0011\rangle$
\end{itemize}

Each transition is optimized independently, and the resulting dynamics are characterized through both direct fidelity evolution and spectral validation against the analytically predicted quasi-energy structure. In all transitions, we used the same coupling constants introduced in Eq.\eqref{7_qubit_hamiltonian}: $g_i = [0.497, 0.499, 0.501, 0.502, 0.500, 0.490]\text{ GHz}$.

It is worth noting that the optimal driving parameters $(A, \sigma, T)$ obtained by CMA-ES exhibit substantial quantitative variations across the target transitions ($A \sim 0.086 - 24.5\,\text{GHz}$, $\sigma \sim 0.16 - 2.0\,\text{ns}$). This heterogeneity originates from the complex, non-convex Floquet landscape, where high-fidelity transfer can be mediated by distinct control regimes—ranging from high-amplitude impulsive kicks driving strong Floquet resonances to broader, lower-amplitude modulations. Rather than locking onto hyper-sensitive, ultra-narrow global optima, the stochastic optimization naturally favors broader resonance basins, yielding working points that are inherently more resilient to parameter variations and static disorder (quantified in Sec.~\ref{sec:robustness}).

Similarly, operating at a small number of stroboscopic cycles ($n_{\text{best}} \in \{2, 4, 6\}$) does not compromise the validity of the Floquet framework: the quasi-energy spectrum and Floquet modes are intrinsic properties of the single-period operator $\widehat{U}(T,0)$. Achieving targeted transfer within few cycles highlights the strength of the engineered resonance while offering two key physical advantages. First, keeping $t_{\text{gate}} = n_{\text{best}} T \ll T_1, T_\Phi$ minimizes exposure to environmental decoherence. Second, it halts the coherent accumulation of calibration errors and static disorder, which inherently scales as $(\widehat{U} + \delta \widehat{U})^n$. The sharp, well-resolved dominant peaks in the FFT spectra (Figs.~\ref{fig:transition10100101}--\ref{fig:transition11000011}) indicate that the dynamics remain well-described by the underlying Floquet quasi-energy structure even in this ultra-fast regime.

\subsection{Transition $|1010\rangle \to |0101\rangle$}
This transition implements a simultaneous exchange of two excitations occupying alternating computational sites into their complementary positions. In terms of the physical chain, the excitations on $Q_1$ and $Q_3$ are transferred to $Q_2$ and $Q_4$ respectively, a process that requires the simultaneous action of interaction pathways mediated by all three buses simultaneously. The dense quasi-energy landscape of the extended architecture makes this a test of the protocol's selectivity. For clarity, we relabel the seven site frequencies $\omega_i$,$i=1,\ldots,7$ of Eq.\eqref{7_qubit_hamiltonian} as $\omega_1,\ldots,\omega_4$ for the four computational qubits and $\omega_{C_1},\omega_{C_2},\omega_{C_3}$ for the three tunable couplers, following the chain ordering $	Q_1-Q_{C_1}-Q_2-Q_{C_2}-Q_3-Q_{C_3}-Q_4$. The optimized system parameters are reported in Table \ref{tab:parameters 1010 0101}.

\begin{figure}[h]
	\centering
	
	\begin{subfigure}{0.98\linewidth}
		\centering
		\includegraphics[width=\linewidth]{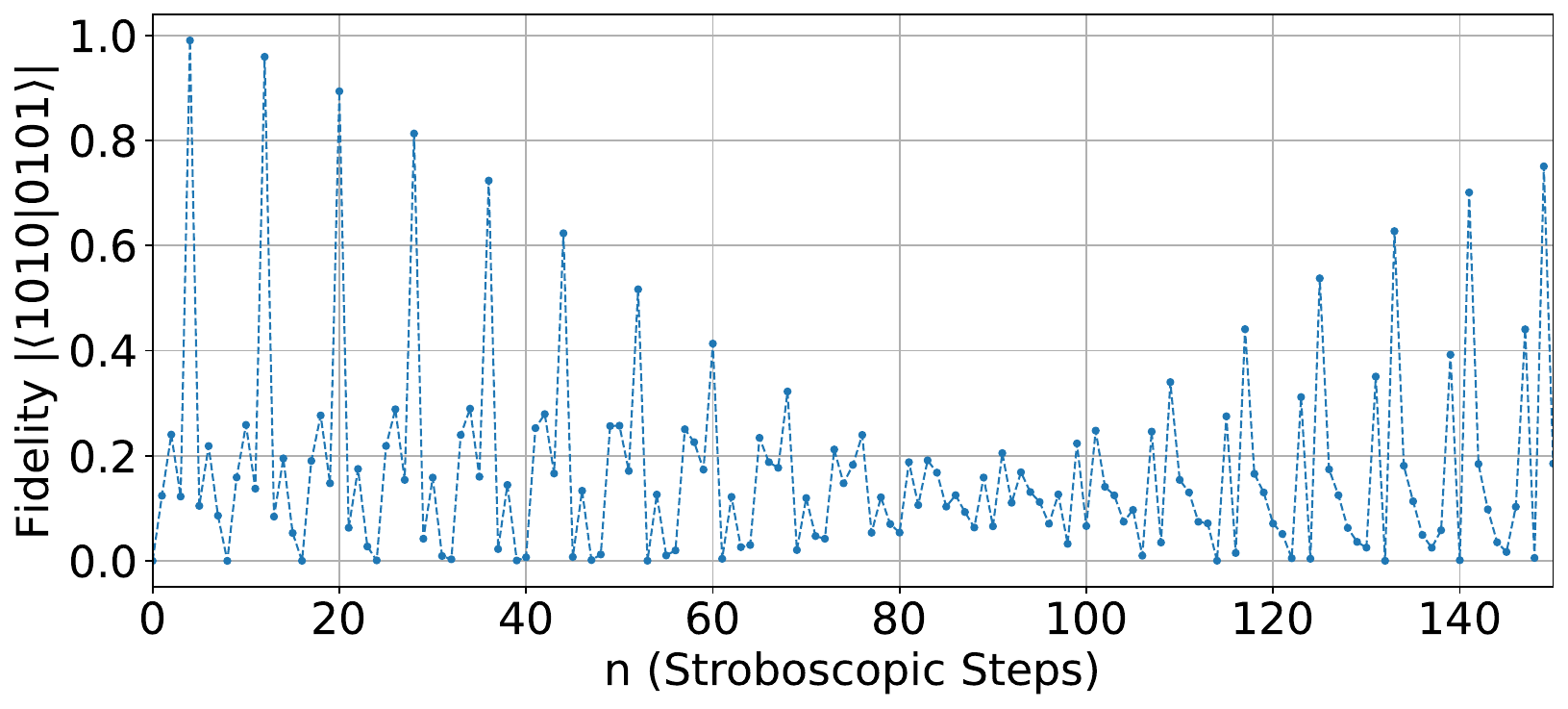}
	\end{subfigure}
	
	\vspace{0.4cm}
	
	\begin{subfigure}{0.98\linewidth}
		\centering
		\includegraphics[width=\linewidth]{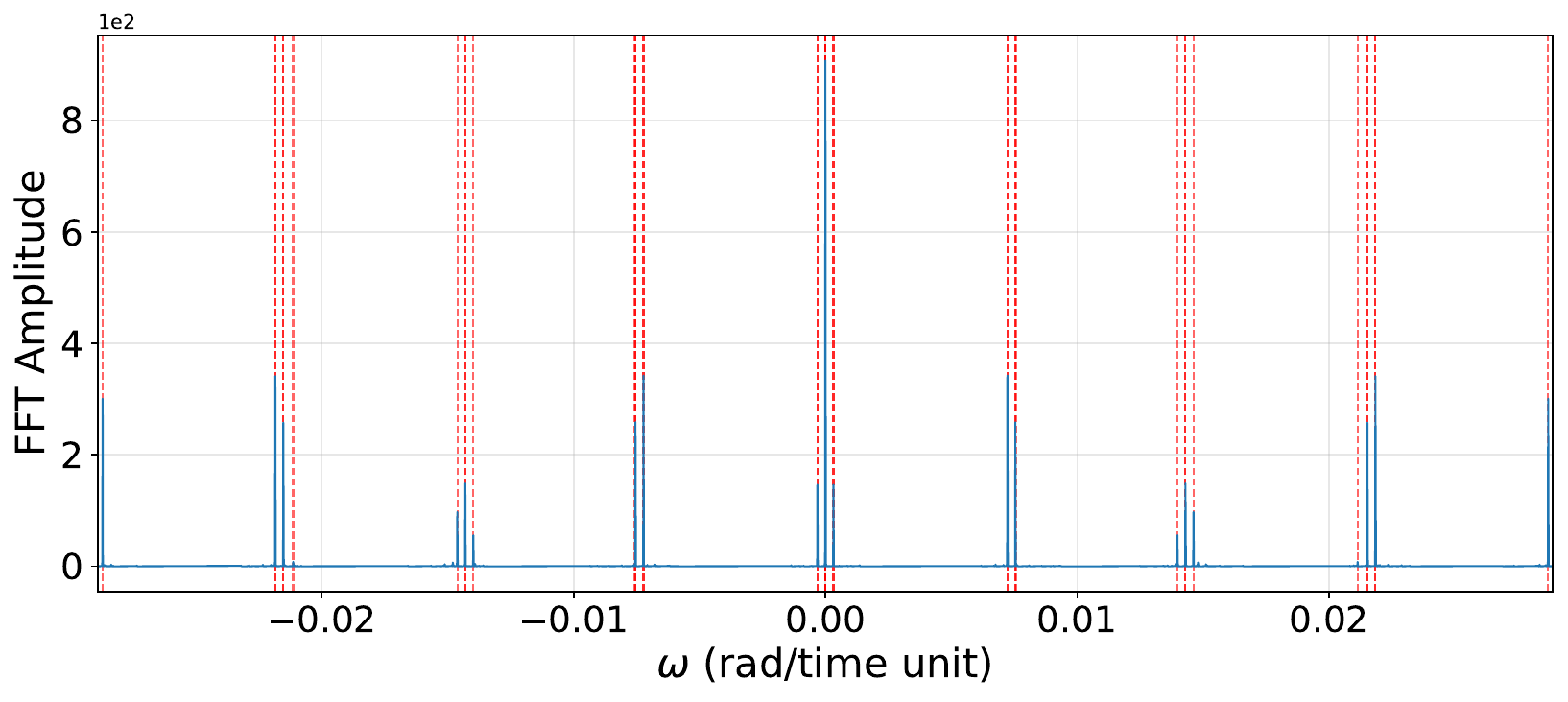}
	\end{subfigure}
	
	\caption{Optimized Floquet dynamics for the transition 
		$|1010\rangle \to |0101\rangle$. (a) Stroboscopic fidelity 
		evolution. (b) FFT spectrum of the fidelity time series; 
		dashed lines indicate the analytically predicted 
		quasi-energy transitions. }
	
	\label{fig:transition10100101}
\end{figure}

\begin{table}[h]
	\centering
	\begin{tabular}{l l}
		\hline
		\textbf{Parameter} & \textbf{Value} \\
		\hline
		$\omega_1, \omega_2, \omega_3, \omega_4$ & $5.1079 ,5.1500, 5.1504, 5.1081$ $GHz$  \\
		$\omega_{C_1},\omega_{C_2},\omega_{C_3}$ & $7.9519, 7.9040, 7.9494$ $GHz$  \\
		$A$ & $24.52$ $GHz$ \\
		$T$ & $108.76$ $ns$\\
		$\sigma$ & $0.1623$ $ns$ \\
		$n_{\text{best}}$ & $4$ \\
		\hline
	\end{tabular}
	\caption{\begin{small}
			System parameters used for numerical simulations of $|1010\rangle \to |0101\rangle$.
	\end{small}}
	\label{tab:parameters 1010 0101}
\end{table}

The fidelity evolution in Fig.\ref{fig:transition10100101}(a) reaches $\mathcal{F} \simeq 0.9902$, with negligible leakage into competing states throughout the evolution. The FFT spectrum in Fig.~\ref{fig:transition10100101}(b) displays a sharp dominant peak in precise agreement with the analytically predicted quasi-energy transition, confirming that the transfer is governed by a single Floquet resonance of order $m$ \cite{resonance, sambe1973steady}. The suppression of sideband amplitudes shows the spectral selectivity of the global driving protocol.

\subsection{Transition $|1001\rangle \to |0110\rangle$}
This transition involves a correlated inward redistribution of two excitations: the excitations initially localized on the outermost computational qubits $Q_1$ and $Q_4$ are coherently transferred to the interior sites $Q_2$ and $Q_3$. Unlike the previous case, it is not an exchange between complementary configurations but a compression of the excitation manifold toward the bulk of the chain. This process requires the coordination of coupling channels spanning the full architecture while suppressing destructive interference from the numerous competing pathways connecting edge and interior sites. The optimized parameters are reported in Table \ref{tab:parameters10010110}.

\begin{figure}[h]
	\centering
	
	\begin{subfigure}{0.98\linewidth}
		\centering
		\includegraphics[width=\linewidth]{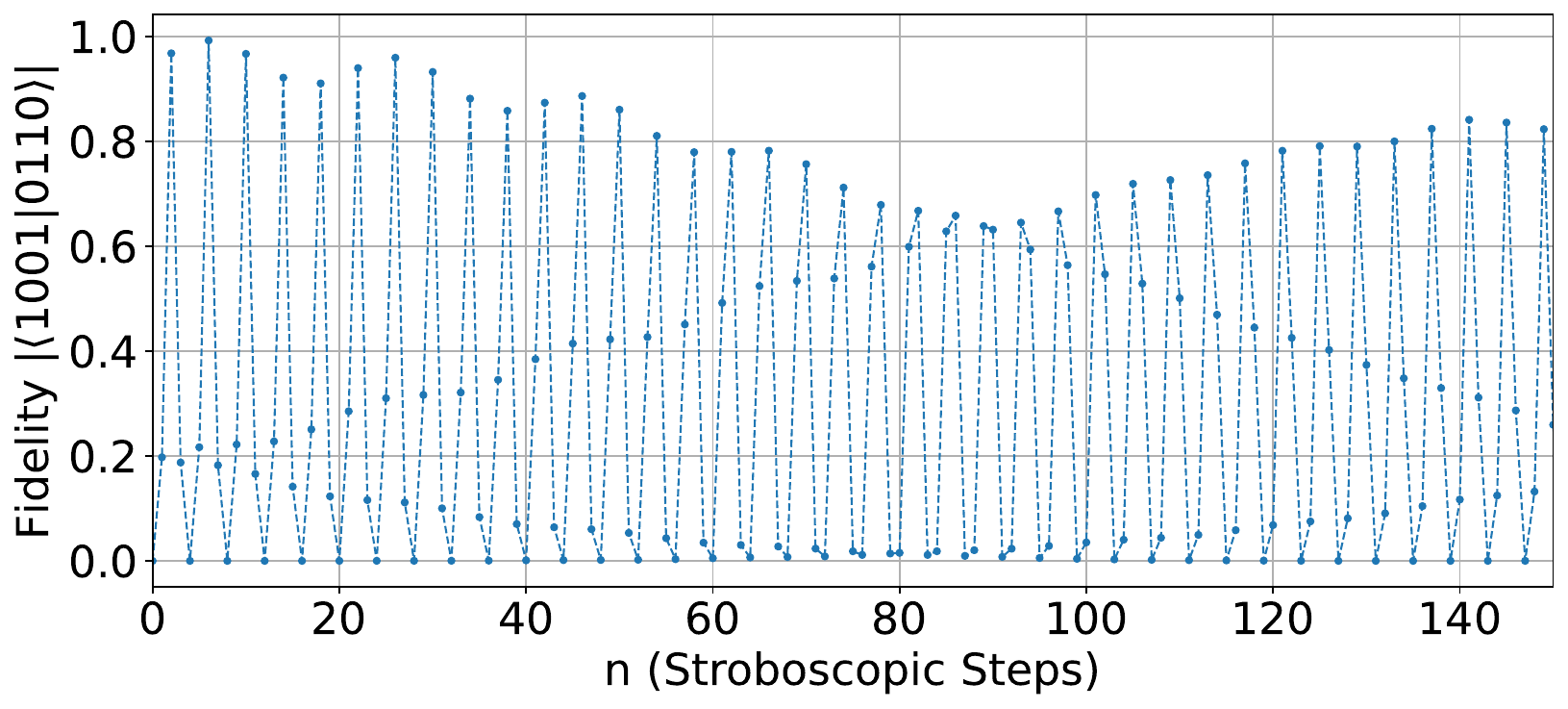}
	\end{subfigure}
	
	\vspace{0.4cm}
	
	\begin{subfigure}{0.98\linewidth}
		\centering
		\includegraphics[width=\linewidth]{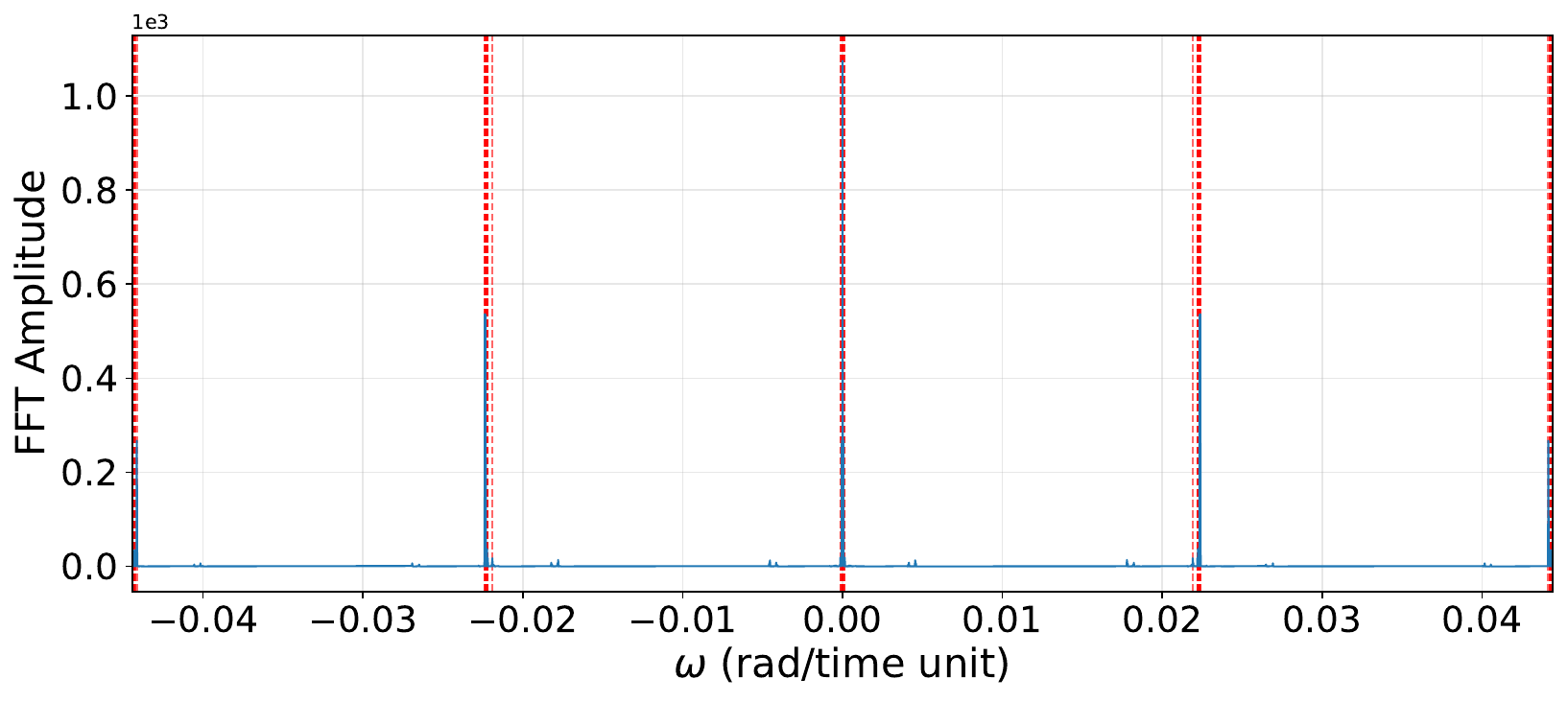}
	\end{subfigure}
	
	\caption{Optimized Floquet dynamics for the transition 
		$|1001\rangle \to |0110\rangle$. (a) Stroboscopic fidelity 
		evolution. (b) FFT spectrum of the fidelity time series; 
		dashed lines indicate the analytically predicted 
		quasi-energy transitions.}
	
	\label{fig:transition10010110}
\end{figure}

\begin{table}[h]
	\centering
	\begin{tabular}{l l}
		\hline
		\textbf{Parameter} & \textbf{Value} \\
		\hline
		$\omega_1, \omega_2, \omega_3, \omega_4$ & $4.5260 ,4.5463, 4.0540, 4.0123$ $GHz$  \\
		$\omega_{C_1},\omega_{C_2},\omega_{C_3}$ & $8.5953, 8.9760, 7.0383$ $GHz$ \\
		$A$ & $0.6643$ $GHz$ \\
		$T$ & $70.72$ $ns$\\
		$\sigma$ & $0.1819$ $ns$ \\
		$n_{\text{best}}$ & $6$ \\
		\hline
	\end{tabular}
	\caption{\begin{small}
			System parameters used for numerical simulations of $|1001\rangle \to |0110\rangle$.
	\end{small}}
	\label{tab:parameters10010110}
\end{table}

The fidelity evolution in Fig \ref{fig:transition10010110}(a) reaches $\mathcal{F} \simeq 0.9931$. The FFT spectrum in Fig \ref{fig:transition10010110}(b) shows a well resolved dominant peak consistent with a Floquet resonance of order $m$ satisfying Eq \eqref{floquet_resonance}, with competing transitions effectively suppressed by the optimized driving parameters \cite{mundada2019suppression, werninghaus2021leakage}. 

\subsection{Transition $|1100\rangle \to |0011\rangle$}

This transition is structured as follows: two excitations co-localized on $Q_1$ and $Q_2$ are coherently transferred to $Q_3$ and $Q_4$, displacing the excitation block across the chain while preserving its internal structure. This can be interpreted as a many-body analogue of directed quantum transport within the computational subspace \cite{bukov2015universal}. Unlike the previous two cases, the initial and final states occupy non-overlapping regions of the chain, so no single local coupling channel can directly connect them, the transfer necessarily proceed through a coordinated multi-channel resonance involving all three buses, making this configuration the most sensitive to spectral crowding and parasitic couplings. The optimized parameters are reported in Table \ref{tab:parameters11000011}.

\begin{figure}[h]
	\centering
	
	\begin{subfigure}{0.98\linewidth}
		\centering
		\includegraphics[width=\linewidth]{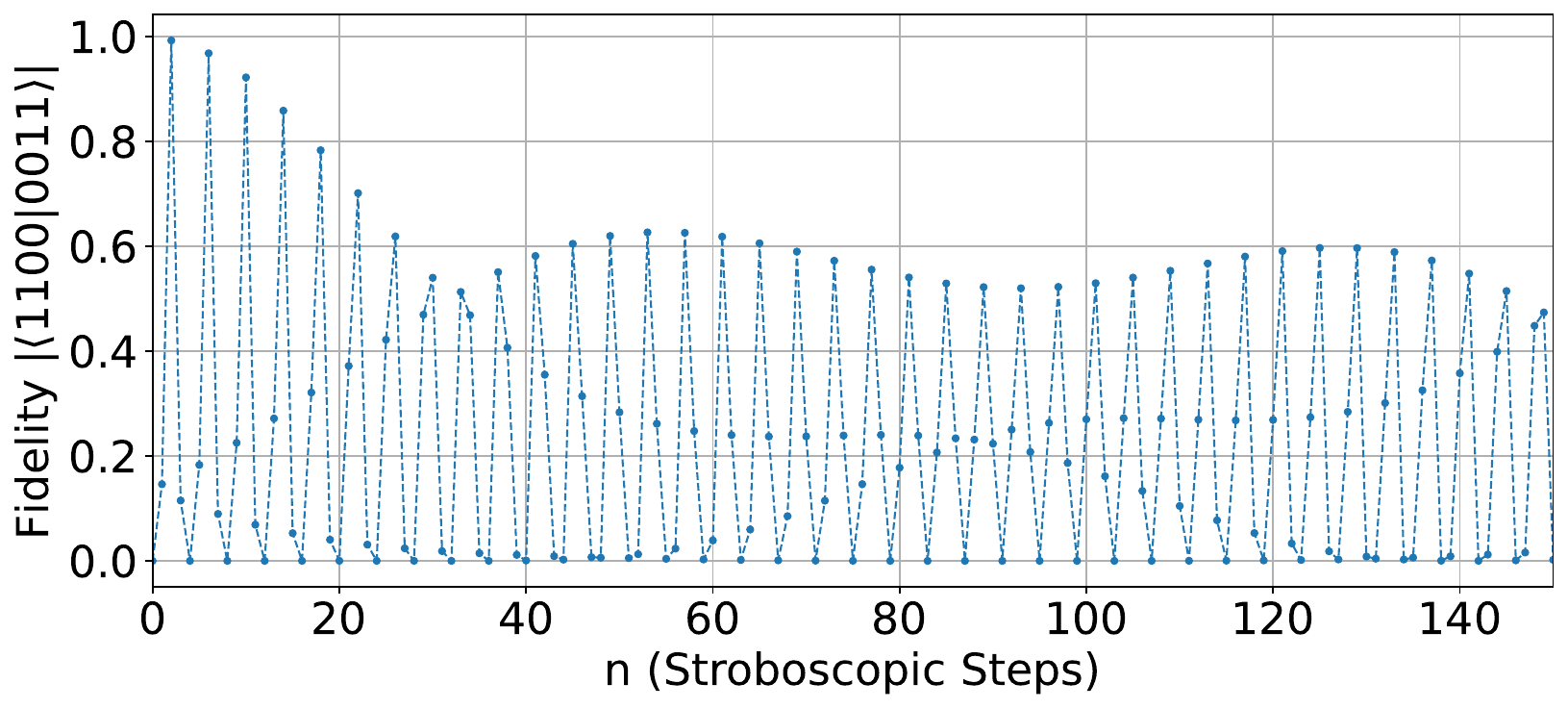}
	\end{subfigure}
	
	\vspace{0.4cm}
	
	\begin{subfigure}{0.98\linewidth}
		\centering
		\includegraphics[width=\linewidth]{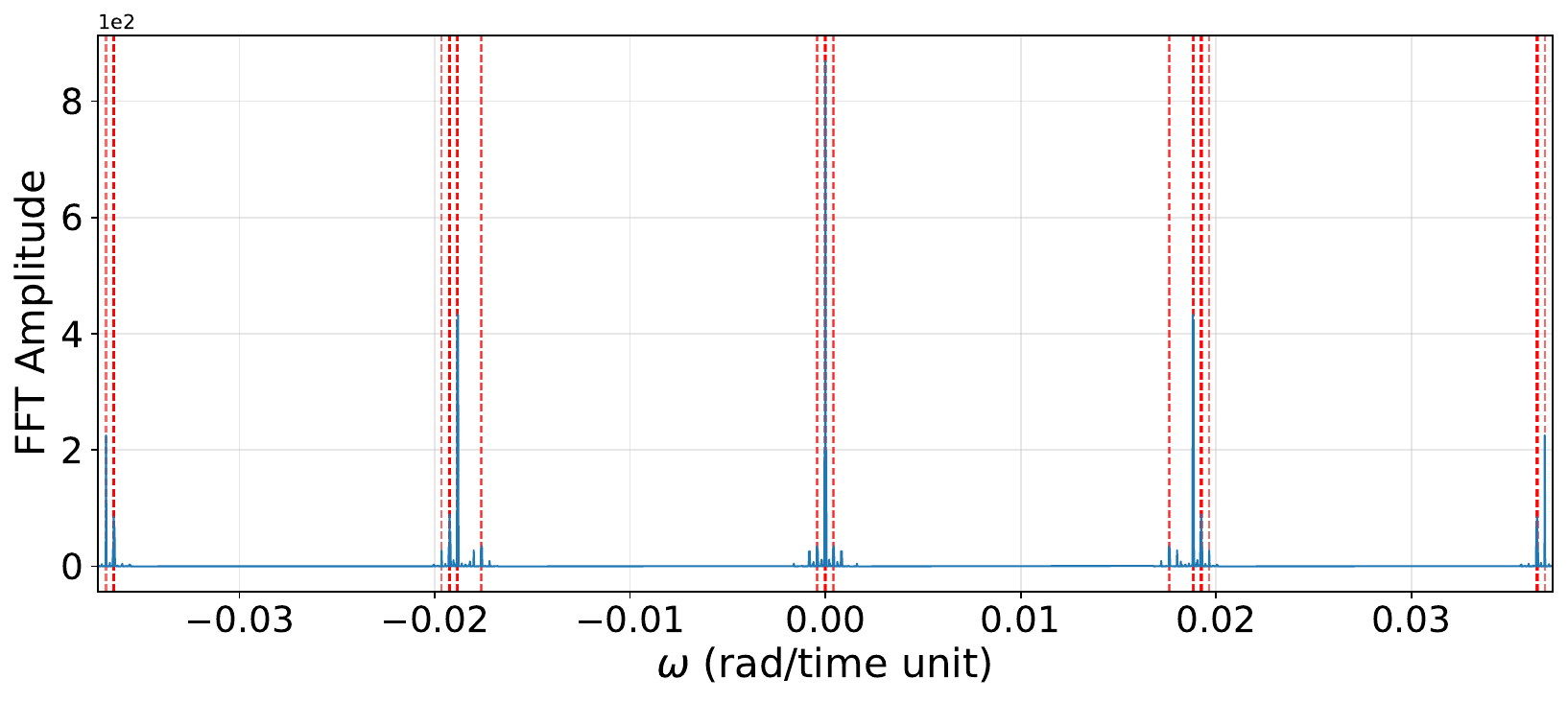}
	\end{subfigure}
	
	\caption{Optimized Floquet dynamics for the transition 
		$|1100\rangle \to |0011\rangle$. (a) Stroboscopic fidelity 
		evolution. (b) FFT spectrum of the fidelity time series; 
		dashed lines indicate the analytically predicted 
		quasi-energy transitions.}
	
	\label{fig:transition11000011}
\end{figure}

\begin{table}[h]
	\centering
	\begin{tabular}{l l}
		\hline
		\textbf{Parameter} & \textbf{Value} \\
		\hline
		$\omega_1, \omega_2, \omega_3, \omega_4$ & $5.5272, 5.5188, 5.5550, 5.4892$ $GHz$\\
		$\omega_{C_1},\omega_{C_2},\omega_{C_3}$ & $7.0014, 8.9175, 7.0006$ $GHz$\\
		$A$ & $0.0862$ $GHz$ \\
		$T$ & $84.36$ $ns$ \\
		$\sigma$ & $2.0$ $ns$\\
		$n_{\text{best}}$ & $2$ \\
		\hline
	\end{tabular}
	\caption{\begin{small}
			System parameters used for numerical simulations  of $|1100\rangle \to |0011\rangle$.
	\end{small}}
	\label{tab:parameters11000011}
\end{table}

As shown in Fig. \ref{fig:transition11000011}(a), the fidelity reaches $\mathcal{F} \simeq 0.9920$. The FFT spectrum in Fig \ref{fig:transition11000011}(b) confirms that the dynamics are dominated by an isolated resonance of order $m$, with the remaining spectral weight strongly suppressed. Despite the structural complexity of this operation, the optimization isolates a dominant resonant manifold, demonstrating that globally driven Floquet architectures can sustain high-fidelity correlated state transfer even when initial and final configurations occupy entirely disjoint regions of the computational subspace.

\section{Sensitivity to Static Parameter Fluctuations}
\label{sec:robustness}

To validate the experimental feasibility of the proposed Floquet-engineered state transfer protocol, it is essential to quantify its sensitivity against both environmental decoherence and calibration imperfections. In state-of-the-art fixed-frequency transmon architectures with tunable-coupler buses of the type considered in this work \cite{architecture, roth2017analysis}, typical energy-relaxation and dephasing times reported for superconducting quantum processors of this class are $T_1 \sim 20-100\mu\text{s}$ and $T_\Phi \sim 10 - 50 \mu s$ \cite{jurcevic2021demonstration, sung2021realization}. We stress that our scale-separation argument depends only on the dimensionless ratio $t_{\text{gate}}/T_1$: the same conclusion holds qualitatively for any platform, provided $t_\text{gate} \ll T_1,T_\Phi$. \cite{krantz2019quantum, burnett2019decoherence, place2021new, tuokkola2025methods, somoroff2023millisecond}. 

In our framework, the co-optimization strategy yields exceptionally short gate durations across all targeted configurations: specifically, $t_{\text{gate}} \sim 430\text{ ns}$ for the $|1010\rangle \rightarrow |0101\rangle$ channel, $t_{\text{gate}} \sim 420\text{ ns}$ for the $|1001\rangle \rightarrow |0110\rangle$ channel, and an ultra-fast $t_{\text{gate}} \sim 170\text{ ns}$ for the long-range $|1100\rangle \rightarrow |0011\rangle$ transition. Comparing these operational times with the decoherence windows highlights a clear scale separation where $t_{\text{gate}} \ll T_\phi \le T_1$. Even under conservative hardware estimates ($T_\phi \sim 10\,\mu\text{s}$), we estimate that the upper bound for the probability of a phase-damping or relaxation event during the multi-excitation gate is $1 - e^{-t_{\text{gate}}/T_\phi} \lesssim 4\%$. This order-of-magnitude estimate suggests that environmental decay may not be the main limiting factor during a single stroboscopic cycle; however, a comprehensive open-system simulation (e.g., via the Lindblad master equation) remains to be performed to fully quantify dissipative and non-Markovian dynamics.

\begin{figure*}[t!]
	\centering
	\makebox[\textwidth][c]{%
		\includegraphics[width=0.34\textwidth]{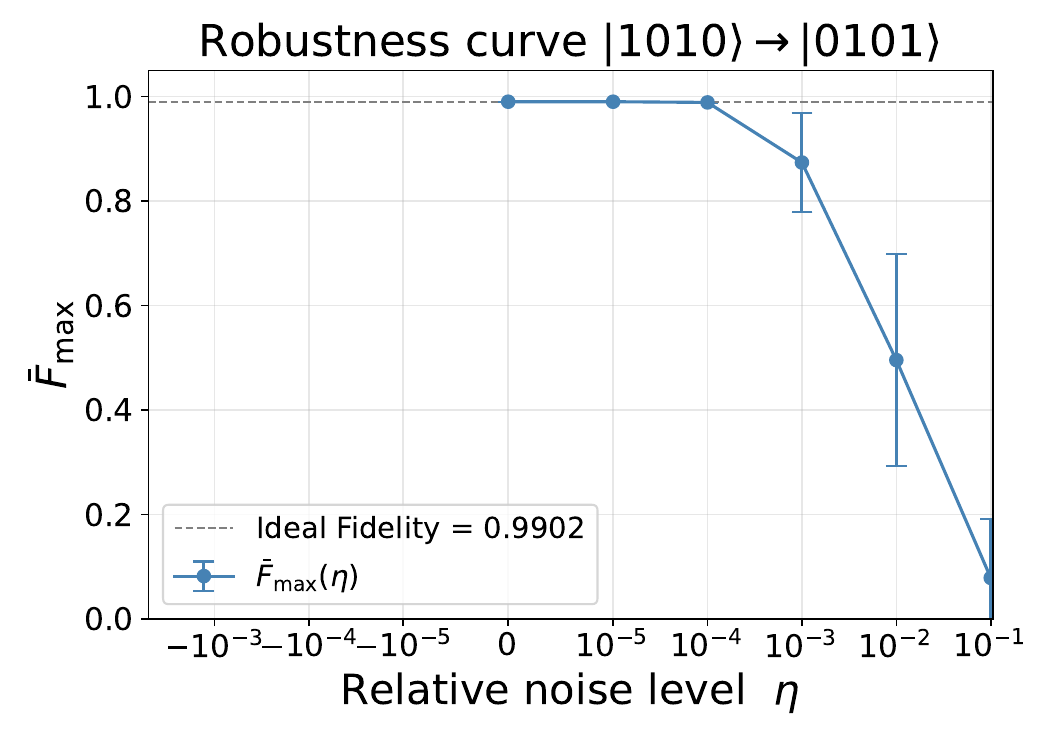}\hfill
		\includegraphics[width=0.34\textwidth]{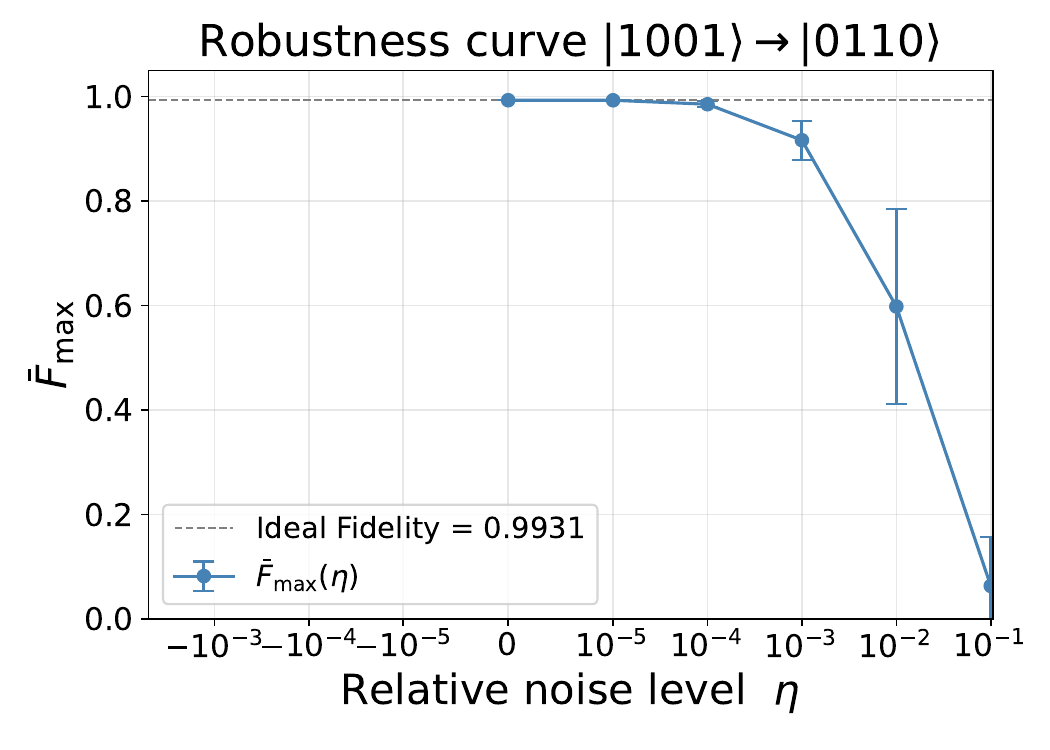}\hfill
		\includegraphics[width=0.34\textwidth]{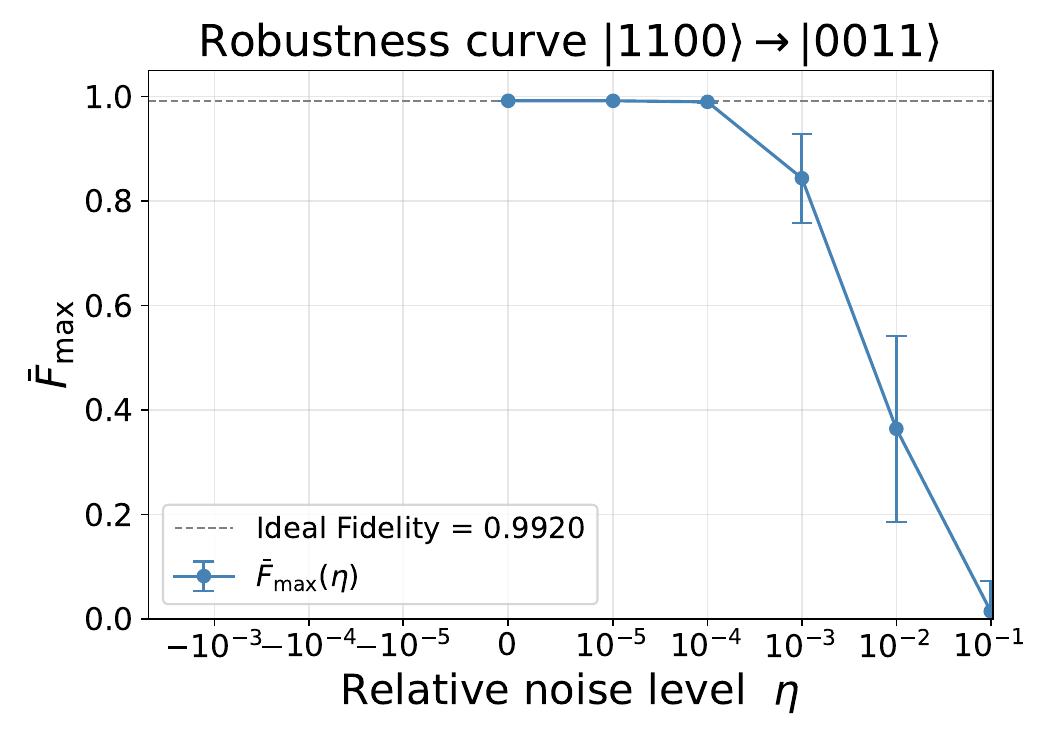}%
	}
	\caption{Ensemble-averaged maximum transfer fidelity $\bar{F}_{\max}$ as a function of the relative static noise level $\eta$ for the three target transitions: (a) $|1010\rangle \rightarrow |0101\rangle$, (b) $|1001\rangle \rightarrow |0110\rangle$, and (c) $|1100\rangle \rightarrow |0011\rangle$. The solid lines correspond to the exact numerical average, error bars represent the standard deviation $\sqrt{\text{Var}}$ within the disorder ensemble, and the shaded regions indicate the statistical spread including shot noise ($N_{\text{shots}}=1000$). The horizontal axis uses a symmetric (signed) logarithmic scale purely for visual spacing around $\eta = 0$; note that $\eta$ is by definition non-negative.}
	\label{fig:robustness_comparison}
\end{figure*}

Conversely, the primary limiting factor for our global resonance mechanism is the presence of static hardware disorders and calibration drifts. Since the state transfer is enabled by a fine-tuned constructive interference between global Floquet quasi-energy modes, structural deviations in the physical parameters can shift the underlying Hamiltonian eigenvalues out of resonance, inducing a loss of contrast in the stroboscopic population dynamics, consistent with previous studies on the detrimental effect of static disorder on coherent state transfer in spin-chain architectures \cite{petrosyan2010state}.

To systematically evaluate this vulnerability, we model a realistic experimental scenario where control parameters dictate clean, unperturbed pulse properties ($A$ and $T$ from the signal generator), whereas the physical architecture parameters suffer from independent Gaussian imperfections. Specifically, the perturbed physical Hamiltonian parameters are drawn according to:
\begin{align}
	\omega_i &\rightarrow \omega_i + \delta\omega_i, \quad \delta\omega_i \sim \mathcal{N}\left(0, \eta^2 \bar{\omega}_{\text{group}}^2\right), \\
	g_i &\rightarrow g_i + \delta g_i, \quad \delta g_i \sim \mathcal{N}\left(0, \eta^2 |g_i|^2\right),
\end{align}
where $\eta$ represents the relative noise scaling factor. The frequency variance is scaled by the average nominal group frequencies of the bare qubits ($\bar{\omega}_q \simeq 4.67\text{ GHz}$) and the tunable couplers ($\bar{\omega}_C \simeq 8.50\text{ GHz}$) respectively, while the disorder in the coupling strengths is applied per-element, keeping the variance proportional to each local nominal coefficient $|g_i|$. We note that restricting the disorder model to Hamiltonian parameters ($\omega_i, g_i$) represents a simplified first-order approximation: explicit control line imperfections, such as pulse amplitude drifts, timing jitter, or finite AWG sampling resolution, are omitted here to isolate the primary impact of static physical calibration errors on the Floquet resonance structure.

We sample the configuration landscape over an ensemble of $N_{\text{noise}} = 50$ independent static disorder realizations for each noise strength $\eta \in [10^{-5}, 10^{-1}]$, tracking the full stroboscopic evolution up to $N_{\max} = n_{\text{best}} + 20$ cycles. To fully characterize the protocol performance across the distinct physical processes introduced in Sec.~\ref{sec: Scaling 7}, this statistical analysis is applied independently to all three target double-excitation transitions: the highly localized $|1010\rangle \rightarrow |0101\rangle$ transfer, the symmetric intermediate-range $|1001\rangle \rightarrow |0110\rangle$ process, and the long-range state exchange $|1100\rangle \rightarrow |0011\rangle$.

Furthermore, to bridge the gap between ideal numerical expectations and real-world experimental readouts, we simulate the effect of quantum projection noise during the state measurement phase. For each disorder realization and at each stroboscopic step $n$, the target state population $F(n)$ is converted into an experimentally accessible estimator $\hat{P}_f(n)$ via a binomial sampling over $N_{\text{shots}} = 1000$ simulated projective measurements \cite{itano1993quantum}, defined as $N_f \sim \text{Binomial}(N_{\text{shots}}, F(n))$ with $\hat{P}_f = N_f / N_{\text{shots}}$.

The comparative scaling analysis under static disorder is illustrated in Fig.~\ref{fig:robustness_comparison}. All three protocols exhibit a highly resilient plateau for weak noise levels up to $\eta \sim 10^{-4}$, where the ensemble-averaged fidelity closely matches the ideal values ($F > 0.99$). 

However, as the disorder strength enters the $\eta \in [10^{-3}, 10^{-2}]$ window, a sharp and severe degradation of the transfer efficiency is observed, revealing a heightened vulnerability of the global resonance mechanism to macroscopic hardware imperfections. The localized $|1010\rangle \rightarrow |0101\rangle$ transition (Fig.~\ref{fig:robustness_comparison}a) proves to be the most resilient configuration, maintaining an average fidelity above $0.90$ at $\eta = 10^{-3}$. Conversely, the longer-range, spatially non-local transitions suffer from a much earlier performance collapse, accompanied by a significant expansion of the statistical variance (represented by the widening error bars). 

This structural fragility is deeply rooted in the underlying many-body quasi-energy spectrum discussed in Sec.~\ref{sec: Scaling 7}. In a large 7-qubit linear architecture, multi-excitation transport relies on high-order, narrow Floquet pathways mediated by a dense manifold of proximate levels. Even sub-percent structural variations ($\eta \sim 10^{-3}$) in the on-site frequencies $\omega_i$ or coupling parameters $g_i$ induce local detunings that alter the destructive interference required to suppress parasitic channels. This triggers unpredictable population leakage across microscopic avoided crossings, breaking down the fine-tuned global resonance condition.

While this performance drop-off under percent-level calibration errors outlines a stringent constraint for purely linear topologies, this limitation is primarily architectural rather than methodological. The strict boundaries of a linear chain inherently restrict the interaction pathways, forcing the protocol to navigate through a crowded, non-degenerate eigenvalue landscape. This vulnerability could be substantially mitigated by shifting from a linear array to a closed ring geometry (periodic boundary conditions). A cyclic architecture naturally introduces spatial symmetries and topological degeneracies that can structurally protect the driving quasi-energies against local parameter fluctuations, significantly widening the high-fidelity operational window. The systematic deployment and optimization of global Floquet engineering within such closed-loop topologies will be the subject of a forthcoming study.

Finally, the simulated shot-noise analysis (shaded bands in Fig.~\ref{fig:robustness_comparison}) indicates that projective measurement fluctuations introduce only a narrow, predictable statistical baseline. Because this readout noise remains significantly smaller than the systematic drop induced by the structural disorder, it does not obscure the underlying physical degradation. This confirms that the exact thresholds of the high-fidelity operational window—and its subsequent breakdown—remain sharply resolvable and characterizable under realistic experimental readout conditions.\\

\section{Conclusions}
\label{sec:conclusion}

In this work, we have investigated the implementation of quantum gates and state transfer protocols in superconducting architectures through Floquet engineering. By initially modelling the external modulation as a periodic train of delta-pulses acting on the longitudinal degree of freedom of flux-tunable transmon buses, we have developed a rigorous stroboscopic framework that maps the gate synthesis problem onto the identification of resonant quasi-energy conditions in the driven system.

In the three-qubit architecture, the combination of analytical resonance estimates, derived from the Floquet spectral decomposition and the BCH expansion, with the CMA-ES optimization strategy has enabled the high-fidelity synthesis of the iSWAP gate.Fourier analysis of the fidelity dynamics confirms that the state transfer is driven by isolated Floquet resonances of well-defined order $m$, while competing channels are effectively suppressed by the pulsed driving protocol.

Extending the framework to a seven-site linear chain of four computational qubits and three tunable buses, we have departed from idealized delta-kicks by implementing a physically realistic periodic train of finite-width Gaussian pulses. We have demonstrated that this global driving protocol can realize qualitatively distinct multi-excitation transfer processes—localized alternation-site exchange ($t_{\text{gate}} \sim 430\text{ ns}$), intermediate-range symmetric redistribution ($t_{\text{gate}} \sim 420\text{ ns}$), and long-range end-to-end block transport ($t_{\text{gate}} \sim 170\text{ ns}$)—all with fidelity approaching unity. The short duration of these gates achieves a crucial scale separation with typical transmon relaxation ($T_1$) and dephasing ($T_\phi$) times, ensuring that open-system dissipative dynamics are effectively negligible during the operation.

Furthermore, we have provided a comprehensive robustness analysis of the protocol under realistic hardware imperfections. While all configurations exhibit a resilient high-fidelity plateau under weak noise ($\eta \le 10^{-4}$), they experience a sharp and severe performance degradation as static parameter disorder reaches the sub-percent level ($\eta \sim 10^{-3}$). This structural fragility is deeply rooted in the dense quasi-energy landscape of the 7-qubit linear architecture, where local detunings break the constructive interference pathways and induce parasitic population leakage across microscopic avoided crossings. Importantly, our simulated shot-noise analysis verified that quantum projection noise introduces only a predictable baseline that does not obscure this underlying physical degradation.

Ultimately, this drop-off underscores an architectural limitation rather than a methodological flaw. To mitigate the spectral crowding of linear topologies, a promising extension involves closed-loop ring geometries with periodic boundary conditions. The inherent spatial symmetries of cyclic architectures can help preserve quasi-energy degeneracies and reduce sensitivity to local parameter fluctuations, thereby broadening the high-fidelity operational window. Investigating global Floquet control in such closed-loop configurations remains an attractive avenue for developing scalable, hardware-efficient quantum routers.

\clearpage
\appendix

\section{General Foundations of Floquet Theory}
\label{app:floquet_foundations}

Floquet theory provides a framework for treating quantum systems governed by time-periodic Hamiltonians, $\widehat{H}(t) = \widehat{H}(t+T)$, where $T$ is the driving period \cite{bukov2015universal, sambe1973steady}. The evolution of a quantum state $|\psi(t)\rangle$ is dictated by the time-dependent Schrödinger equation:

\begin{equation}
	i\frac{d}{dt}|\psi(t)\rangle = \widehat{H}(t)|\psi(t)\rangle.
\end{equation}

\begin{figure}[b]
	\centering
	\begin{tikzpicture}[scale=0.6, transform shape]
		
		\draw[thick,->](-3,0) -- (10,0) node[anchor=north west]{time};
		\draw[thick, domain=-2.5:8.5, samples=100]
		plot (\x,{1.5*sin(\x*180/pi*2)});
		\node[circle,fill=black, inner sep=1pt](t1) at (-1,0){};
		\node[below, xshift=-2mm] at (t1) {$t_1$};
		\node[circle, fill=black, inner sep=1pt] (t2) at (7,0){};
		\node[below, xshift=-2mm] at (t2) {$t_2$};
		\draw[thick, dashed] (-1,0)--(-1,-2.5);
		\draw[thick, dashed] (7,0)--(7,-2.5);
		\draw[thick, <->] (-1,-2.5)--(7,-2.5);
		\node at (3,-2.2) {$t$}; 
		\node[circle,fill=red, inner sep=1pt](t0) at (0,0){};
		\node[above, xshift=-2mm, text=red] at (t0) {$t_0$};
		\node[circle, fill=red, inner sep=1pt] (t0') at (6.28 ,0){};
		\node[above, xshift=-2mm,text=red] at (t0') {$t_0'$};
		\draw[thick, dashed, color=red] (0,0)--(0,2.5);
		\draw[thick, dashed, color=red] (6.28,0)--(6.28,2.5);
		\draw[thick, color=red, <->] (0,2.5)--(6.28,2.5);
		\node[text=red] at (3.14,2.7) {$nT$};
		
	\end{tikzpicture}
	\caption{\begin{small} Floquet gauge: The system evolves from time $t_1$ to time $t_2$. The stroboscopic evolution starts at time $t_0$ which can be chosen to be anywhere within the first period $[t_1,t_1+T)$. The choice of the $t_0$ (Floquet gauge) in general affects the form of the stroboscopic Floquet Hamiltonian $\widehat{H}_F[t_0]$. \end{small}}
	\label{fig:FloquetGauge}
\end{figure}
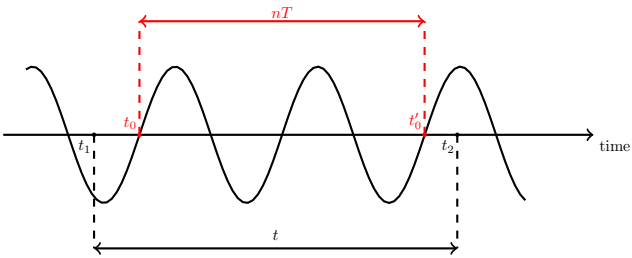

The dynamics of a periodic system is captured by the unitary time-evolution operator $\widehat{U}(t,t_0)$. According to the Floquet theorem, the propagator for any integer number of periods $n$ can be decomposed as:

\begin{equation}
	\widehat{U}(t+nT,t_0) \equiv \widehat{U}(t,t_0)[\widehat{U}(t_0+T,t_0)]^n,
	\label{floquet_theory}
\end{equation}

Thus, the long-time behaviour is entirely determined by the one-period evolution operator, or Floquet propagator:

\begin{equation}
	\widehat{F}_0 \equiv \widehat{U}(t_0+T,t_0) = e^{-i\widehat{H}_F[t_0]T},
\end{equation}

where $\widehat{H}_F[t_0]$ is the Hermitian Floquet Hamiltonian. Its diagonalization yields the Floquet modes $|u_j(t_0)\rangle$ and their associated quasi-energies $\epsilon_j$:

\begin{equation}
	\widehat{F}_0|u_j(t_0)\rangle = e^{-i\epsilon_j T}|u_j(t_0)\rangle,
\end{equation}

with the general solution $|\psi_j(t)\rangle = e^{i\epsilon(t-t_0)}|u_j(t)\rangle$, where $|u_j(t)\rangle$ is periodic in $T$. The quasi-energies are defined modulo $\Omega$ (with $\Omega = 2\pi/T$), in direct analogy with the quasi-momenta of Bloch's theorem for spatially periodic system \cite{condensed_matter}, an important analogy in the exploration of conditions of resonance within the work.

The dependence of $\widehat{H}_F$ on the initial time $t_0$ is referred to as the Floquet gauge \cite{bukov2015universal}. Although the explicit from of $\widehat{H}_F[t_0]$ depends on this choice, different gauges are related by unitary transformations,

\begin{equation}
	\widehat{H}_F[t_1] = \widehat{U}(t_1,t_0)\,\widehat{H}_F[t_0]\,\widehat{U}^\dagger(t_1,t_0),
\end{equation}

so that the quasi-energy spectrum is gauge invariant while the Floquet modes transform covariantly. Physical observables evaluated stroboscopically therefore remain independent of the specific choice of $t_0$ (see Fig. \ref{fig:FloquetGauge}), and the full propagator between arbitrary times $t_1$ and $t_2$ can always be reconstructed as:

\begin{equation}
	\widehat{U}(t_2,t_1) = \widehat{U}(t_2,t_0+nT)e^{-i\widehat{H}_F[t_0]nT}\widehat{U}(t_0,t_1).
\end{equation}

\section{Floquet Resonance Condition and Perturbative Estimate}

\subsection{General Floquet Resonance Condition} \label{app:Resonance}

The resonance condition is a direct consequence of the periodic structure of the Floquet quasi-energy spectrum. 
To induce a coherent transition between two Floquet modes, the driving parameters must be tuned such that a quasi-energy degeneracy occurs modulo the driving frequency. This condition reads

\begin{equation}
	\epsilon_\alpha^{(m)} = \epsilon_\beta^{(n)},
\end{equation}

where $\epsilon_{\alpha}^{(m)} = \epsilon_\alpha + m\Omega$ denotes the $m$-th Floquet replica of the quasi-energy $\epsilon_\alpha$. Since quasi-energies are defined modulo $\Omega$, the degeneracy condition can be rewritten as

\begin{equation}
	\epsilon_\alpha + m\Omega = \epsilon_\beta + n\Omega,
\end{equation}

which implies

\begin{equation}
	\Delta \epsilon_{\alpha\beta} = (n-m)\Omega \equiv k\Omega,
	\quad k \in \mathbb{Z}.
\end{equation}

This expression represents the general Floquet resonance condition. It states that resonant transitions occur when the quasi-energy difference between two modes equals an integer multiple of the driving frequency. Physically, this corresponds to the exchange of $k$ energy quanta $\Omega$ between the system and the periodic drive.

\subsection{Analytical estimates via Floquet spectral decomposition} \label{app:estimate}

The fidelity can be evaluated through a Floquet spectral decomposition by performing a basis transformation from the computational basis to the Floquet basis, where the resonance condition can be naturally applied. The fidelity is defined as:

\begin{equation}
	\mathcal{F}_{i\to f} = |\langle f |\widehat{U}(nT)|i\rangle |^2
\end{equation}

where $|i\rangle$ and $|f\rangle$ denote, respectively, the initial and final states in the computational basis.

These states can be rewritten in the Floquet basis in the general form:

\begin{equation}
	\begin{split}
		|i\rangle = {} & \sum_\alpha c_\alpha|\alpha\rangle \qquad c_\alpha = \langle \alpha | i\rangle \\
		|f\rangle = {} & \sum_\alpha b_\alpha|\alpha\rangle \qquad b_\alpha = \langle \alpha | f\rangle
	\end{split}
\end{equation}

where $\{|\alpha\rangle\}$ is the set of Floquet eigenstates.

Substituting these expressions into the fidelity definition yields:

\begin{equation}
	\begin{split}
		\mathcal{F}_{i \to f} = {} & |\langle f |\widehat{U}(nT)|i\rangle |^2 \\
		= {} & \left|\sum_{\alpha,\beta} c_\alpha b_\beta^* \langle \beta |\widehat{U}(nT)|\alpha\rangle \right|^2
	\end{split}
\end{equation}

Expressing the evolution operator in the Floquet basis:

\begin{equation}
	\widehat{U}(nT)|\alpha\rangle = e^{-i\epsilon_\alpha nT}|\alpha\rangle
\end{equation}

we obtain:

\begin{equation}
	\begin{split}
		\mathcal{F}_{i\to f} = {} & \left|\sum_{\alpha,\beta} c_\alpha b_\beta^* e^{-i\epsilon_\alpha nT}\langle \beta |\alpha\rangle \right|^2 \\
		= {} & \left|\sum_\alpha c_\alpha b_\alpha^* e^{-i\epsilon_\alpha nT}\right|^2 \\
		= {} & \sum_\alpha |c_\alpha|^2|b_\alpha|^2 + \sum_{\alpha\neq\beta} c_\alpha b_\alpha^* c_\beta^* b_\beta e^{-i(\epsilon_\alpha - \epsilon_\beta)nT}
	\end{split}
\end{equation}

At this stage, the resonance condition $\Delta\epsilon_{\alpha\beta} = m\Omega$ can be applied to the interference terms. In practice, this condition is most relevant not for all possible pairs $(\alpha,\beta)$, but primarily for the dominant weighted transitions contributing most significantly to the fidelity evolution.


\clearpage


\bibliographystyle{unsrtnat} 
\bibliography{bibliography}  

\end{document}